\documentclass[prl,aps,twocolumn,superscriptaddress]{revtex4-1}

\usepackage{graphicx,color}
\usepackage{amsthm}
\usepackage{amsfonts}
\usepackage{algorithmic}
\usepackage{enumerate}
\usepackage{latexsym}
\usepackage{amsmath}
\usepackage{amssymb}
\usepackage[colorlinks=true,citecolor=blue,linkcolor=blue]{hyperref}
\def\avg#1{\left\langle#1\right\rangle}

\begin{document}
\title{Strain-tuning of edge magnetism in zigzag graphene nanoribbons}

\author{Guang Yang}
\affiliation{Department of Physics, Beijing Normal University,
Beijing 100875, China}

\author{Baoyue Li}
\affiliation{School of Physics and Electronic-Electrical Engineering, Ningxia University,
Yinchuan 750021, China}
\author{Wei Zhang}
\affiliation{Department of Physics, Beijing Normal University,
Beijing 100875, China}
\author{Miao Ye}
\affiliation{College of Information Science and Engineering, Guilin University of Technology, Guilin 541004, China}
\author{Tianxing Ma}
\email{txma@bnu.edu.cn}
\affiliation{Department of Physics, Beijing Normal University,
Beijing 100875, China}

\date{\today}
\begin{abstract}
Using the determinant quantum Monte-Carlo method, we elucidate the strain tuning of edge magnetism in zigzag graphene nanoribbons. Our intensive numerical results show that a relatively weak Coulomb interaction may induce a ferromagnetic-like behaviour with a proper strain, and the edge magnetism can be enhanced greatly as the strain along the zigzag edge increases, which provides another way to control graphene magnetism even at room temperature.
\end{abstract}
\maketitle

\section{Introduction}
The possible existence of magnetism in graphene has been pursued intensively since this material was first isolated\cite{Novoselov2004,Novoselov2005}.
In past years, many theoretical proposals have been put forward on inducing magnetism, including the use of strain, carrier doping, atomic defects, grain boundaries, vacancies, hydrogen chemiadsorption, and different shapes of edges or structures\cite{Vianagomes2009Magnetism,Magda2014,Li2014,Han2014,Palacios2008,RevModPhys.81.109,
PhysRevB.79.035405,PhysRevLett.106.226401,Mitsutaka1996,
PhysRevB.72.174431,PhysRevB.92.045426,Yazyev2007,Yazyev2010,PhysRevLett.101.037203,
PhysRevLett.104.096804,PhysRevLett.111.166101,MaAPL2010,PhysRevLett.100.186803,JAPMa2012,*LiJPCM2016,*PhysRevB.91.075410,*PhysRevB.94.075106}.
From a theoretical point of view, electronic correlation is suggested to play a key role in the possible existence
of magnetism in graphene-based materials,
and J. Viana-Gomes $et. al$ predicted that strain could enhance the magnetic order at edges of ribbons and graphene quantum dots from a tight-binding perspective in their pioneer works\cite{Vianagomes2009Magnetism}.
 In perfect graphene, the interplay between Coulomb interactions and the Van Hove singularity in the density of states may lead to strong ferromagnetic fluctuations in heavily doped single layer graphene systems\cite{MaAPL2010}, as well as a possible ferromagnetic solution in biased bilayer graphene\cite{PhysRevLett.100.186803}.
However, realizing a high doping level in graphene-based material is a challenging problem\cite{LiVHS2010}, and all the theoretical proposals are awaiting experimental realization.

\begin{figure}
\includegraphics[scale=0.475]{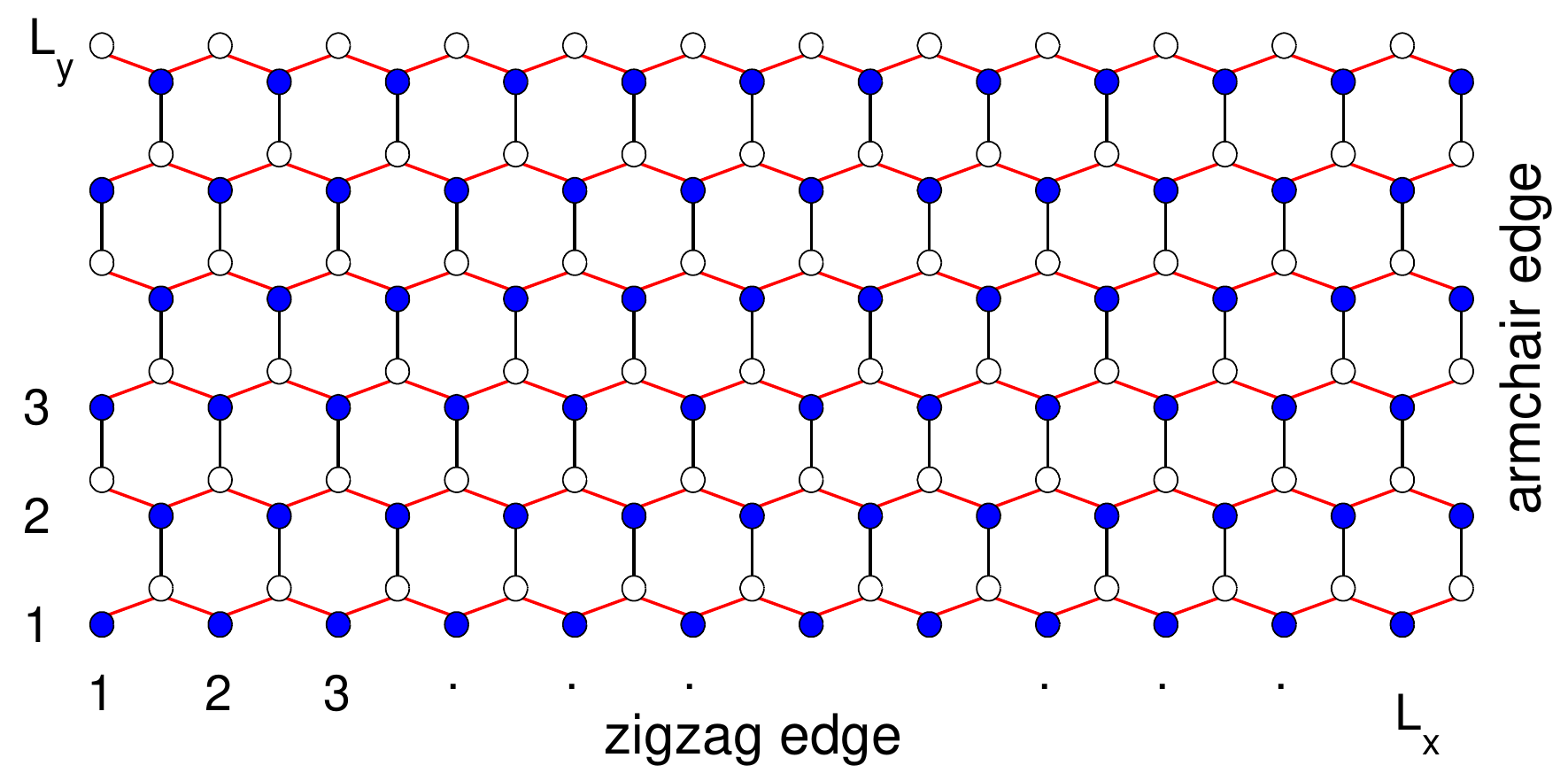}
\caption{(Colour online) A piece of a honeycomb lattice displaying zigzag edges with $L_{y}=6$, which defines the width of the ribbon in the transverse direction, and $L_{x}=12$, indicating the length in the longitudinal direction. The lattice size $N=2\times12\times6=144$, where the blue and white circles indicate the A and B sublattices, respectively. We consider the strain along the zigzag direction. The dark line indicates $t_1=t$; red lines indicate $t_2=t_3=t-\Delta$. Here, $t$ represents the nearest hopping term, and $\Delta$ represents the effect of strain.}
\label{Fig:Sketch}
\end{figure}

In the vast field of carbon-based nanostructures, a central concept is the fact that their physical properties can change dramatically depending on their electronic structure. As for graphene which is hexagonal Bravais lattice, each unit cell consists of two carbon atoms so that there are two sublattice in graphene, and due to this bipartite nature, two kinds of edge can exist: zigzag or armchair, as shown in Fig.\ref{Fig:Sketch}. It is further proposed that the zigzag-shaped edges of the graphene nanoribbons may hold stable magnetism even at room temperature and that their magnetic properties can be controlled by external electric fields\cite{Son2006} or by using hydrogen atoms\cite{Gonzlez2016},which  may open a new path to realize spintronics at the nanometre scale\cite{Magda2014}. Most recently, a bottom-up synthesis of zigzag graphene nanoribbons has been successfully achieved through surface assisted polymerization and cyclodehydrogenation of specifically designed precursor monomers to yield atomically precise zigzag edges\cite{Ruffieux2016}. This provides a great opportunity for scientific advancement in exploring edge magnetism in zigzag graphene nanoribbons. The graphene nanoribbons are much more promising as a way towards the realization of electronic and spintronic devices that can operate at room temperature.

In this paper, we explore a different avenue to control graphene magnetism. Here, we show that the strain along the zigzag edge could induce a ferromagnetic-like behaviour with a relatively weak Coulomb interaction. In addition, our results suggest that edge magnetism with a possible high Curie temperature is enhanced greatly as the strain increases. The system we investigated is shown in Fig.\ref{Fig:Sketch}.

\section{Model and methods}
The system we study is graphene nanoribbons with zigzag edges. In Fig.\ref{Fig:Sketch}, a sketch of a lattice with $L_x=12$ and $L_y=6$ is shown, where the blue and white circles indicate the A and B sublattices respectively. The strain is applied along the zigzag direction, as shown in the figure. Dark lines indicate $t_1=t$ ($\approx 2.7$ eV), and red lines indicate $t_2=t_3=t-\Delta$. Here, $t$ represents the usual nearest hopping term, and $\Delta$ represents the amplitude of the strain.

The Hamiltonian of strained zigzag graphene nanoribbons can be described by the following Hubbard model:
\begin{eqnarray}
H &=&-\sum_{i\eta\sigma
}t_{\eta}(a_{i\sigma }^{\dag }b_{i+\eta\sigma }+h.c.)
+U\sum_{i}(n_{ai\uparrow}n_{ai\downarrow}
+n_{bi\uparrow}n_{bi\downarrow}) \notag \\
&&-\mu \sum_{i\sigma }(n_{ai\sigma }+n_{bi\sigma }) 
\end{eqnarray}
where $t_{\eta}$ is the nearest hopping integral, $\mu$ is the chemical potential and $U$ is the on-site repulsion. Here, $a_{i\sigma}$ ($a_{i\sigma}^{\dagger}$) annihilates (creates) electrons at the site $\mathbf{R}_{i}$ with spin $\sigma$ ($\sigma=\uparrow,\downarrow$) on sublattice A, and $b_{i\sigma}$ ($b_{i\sigma}^{\dagger}$) acts on electrons of sublattice B such that $n_{ai\sigma}=a_{i\sigma}^{\dagger}a_{i\sigma}$ and $n_{bi\sigma}=b_{i\sigma}^{\dagger}b_{i\sigma}$. In the calculation, we apply periodic boundary conditions in the $x$ direction and open boundary conditions at the zigzag edge, and thus, we have a nanoribbon with zigzag edges.

It has been reported that graphene has been established in experiments as the strongest material ever measured\citep{Lee385}. It is able to sustain reversible elastic deformations up to $20\%$ according to both \emph{ab initio} calculations and experiments\cite{PhysRevB.76.064120,Kim2009,Amorim20161}.
In addition, the effect of tensional strain on the electronic properties of graphene has also been proved theoretically through the tight-binding approach and density functional calculations\cite{Ribeiro2009,PhysRevB.80.045401}.
The applied stress changes the band structure of the materials as a consequence of the modification of interatomic distances, which in turn implies a change in the electronic-hopping parameters $t_{\eta}$. The hopping parameter dependence on the strain has been analysed \cite{PhysRevB.80.045401}. According to this, a deformation on the order of $20\%$ corresponds to $\Delta t=0.5t$. To explore the importance of strain on the magnetism of graphene nanoribbons, we study strains in the range of $\Delta =0.0\sim0.50t$. For investigating the electronic correlation-induced edge magnetism theoretically, it has been widely testified that an extended Hubbard model is appropriate for a graphene-based material, and the interaction should be approximately $1.6t$\cite{PhysRevLett.111.036601}.
We investigate the interaction-dependent magnetism, and to emphasize the effect of electronic correlations on the magnetism in such system, we report results associated with $U$ in the range of $1.0t\sim4.0t$. In this region of interaction, the nonperturbative numeric approach used in the present work, the determinant quantum Monte-Carlo (DQMC) method, is a very powerful tool for treating both the edge geometry and the interaction. For more details on this widely used method, we refer the reader to Refs.~\cite{PhysRevD.24.2278,PhysRevB.31.4403,MaAPL2010}.

To explore the magnetic properties of zigzag graphene nanoribbons, one could calculate the magnetic susceptibility by defining the spin susceptibility
in the $z$ direction at zero frequency as
\begin{equation}
\chi(q) =\frac{1}{N_{s}}\int_{0}^{\beta}d\tau\sum_{d,d'=a,b} \sum_{i,j} e^{iq\cdot(i_{d}-{j_{d^{\prime}}})}\langle m_{i_{d}}(\tau)\cdot m_{j_{d^{\prime}}}(0)\rangle \notag
\end{equation}
Here, $m_{ia}(\tau)=e^{H\tau}m_{ia}(0)e^{-H\tau}$ with $m_{ia}=a_{i\uparrow}^{\dagger}a_{i\uparrow}-a_{i\downarrow}^{\dagger}a_{i\downarrow}$ and $m_{ib}=b_{i\uparrow}^{\dagger}b_{i\uparrow}-b_{i\downarrow}^{\dagger}b_{i\downarrow}$.
In present study, we are focusing on the edge magnetism, and in the following,
we investigate the edge magnetic susceptibility $\chi_e$ by making the summation over the
sites on both the top zigzag edge and the bottom zigzag edge, and then taking an average corresponding to the total number of sites $N_{e}=2\times L_x$ at the edges.

\begin{figure}
\centering
\includegraphics[scale=0.5]{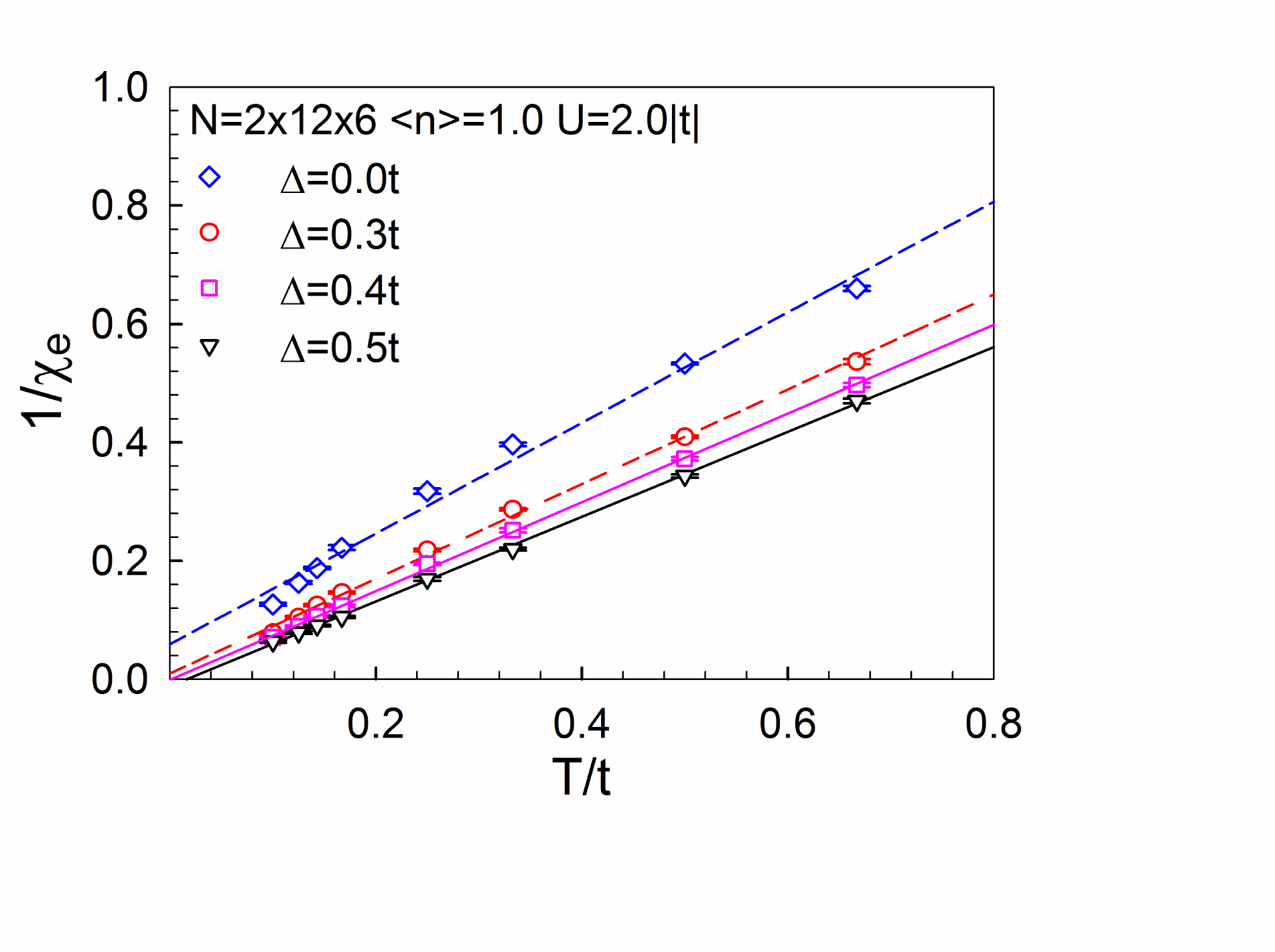}
\caption{(Colour online) The edge magnetic susceptibility at $U=2.0t$ for different strains.}
\label{Fig:2}
\end{figure}

\section{Results and discussions}
The temperature-dependent magnetic susceptibility plays a key role in understanding the behaviour of magnetism. We first show the temperature-dependent edge magnetic susceptibility in Fig.\ref{Fig:2} at half fillings for $U=2.0t$ with $L_{x}=12$ and $L_{y}=6$, in which we try to emphasize the importance of strain on the edge magnetism. In Fig.\ref{Fig:2}, 1/$\chi_{e}(T)$ (symbols) are presented at different strains for $U=2.0t$ and $\avg{n}=$1.0 with linear fittings (dashed lines).
They exhibit the Curie-Weiss law
\begin{equation}
1/\chi_e=(T-T_c)/A,
\label{Tc}
\end{equation}
which describes the magnetic susceptibility for a ferromagnetic material in the temperature region above the Curie temperature.
From Eq. \ref{Tc}, one may see that at $T = T_c$, the $\chi_e$ tends to diverge, and we may estimate the $T_c$ from the fitting
data where the $\chi_e$ may diverge at some temperature. Specifically, as strains are larger than $0.40t$, the intercepts on the $T$ axis are finite, yielding a finite $T_c$. Moreover, the finite $T_c$ increases as the strain increases, which means that one can tune the edge magnetism by changing the stress. For example, at $\Delta=0.5t$, the possible $T_c$ is approximately 0.016$t$ ($\approx$ 400 K) while $T_c$ at $\Delta=0.4t$ is approximately 0.001$t$ ($\approx$25 K) for $U=2.0t$.

There is a visible deviation of the data for the straight behavior, especially that at the low temperature region of Fig.\ref{Fig:2}.
This deviation is mostly caused by the increasing error bars as the temperature decreases. However, according to the origin of Curie-Weiss law, the Curie-Weiss law describes the system at high temperature regime where $T> T_{c}$,  and the data for high temperature have a better linearity. Thus, the estimated critical temperature is basically reliable as which is primarily extrapolated from the high temperature data. We may roughly estimate the error of $T_c$ by using the susceptibility at the lowest temperature as the $\chi_e(T_{lowest})$ should contribute to the deviation mostly. $\delta T_c/T_c=[A\delta \chi_e(T_{lowest})/\chi^2_e(T_{lowest})]/T_c$, which is around 15 percent for the shown data,  indicating that
 the value of $T_c$ should be statistically distinguishable from zero.

\begin{figure}
\centering
\includegraphics[scale=0.5]{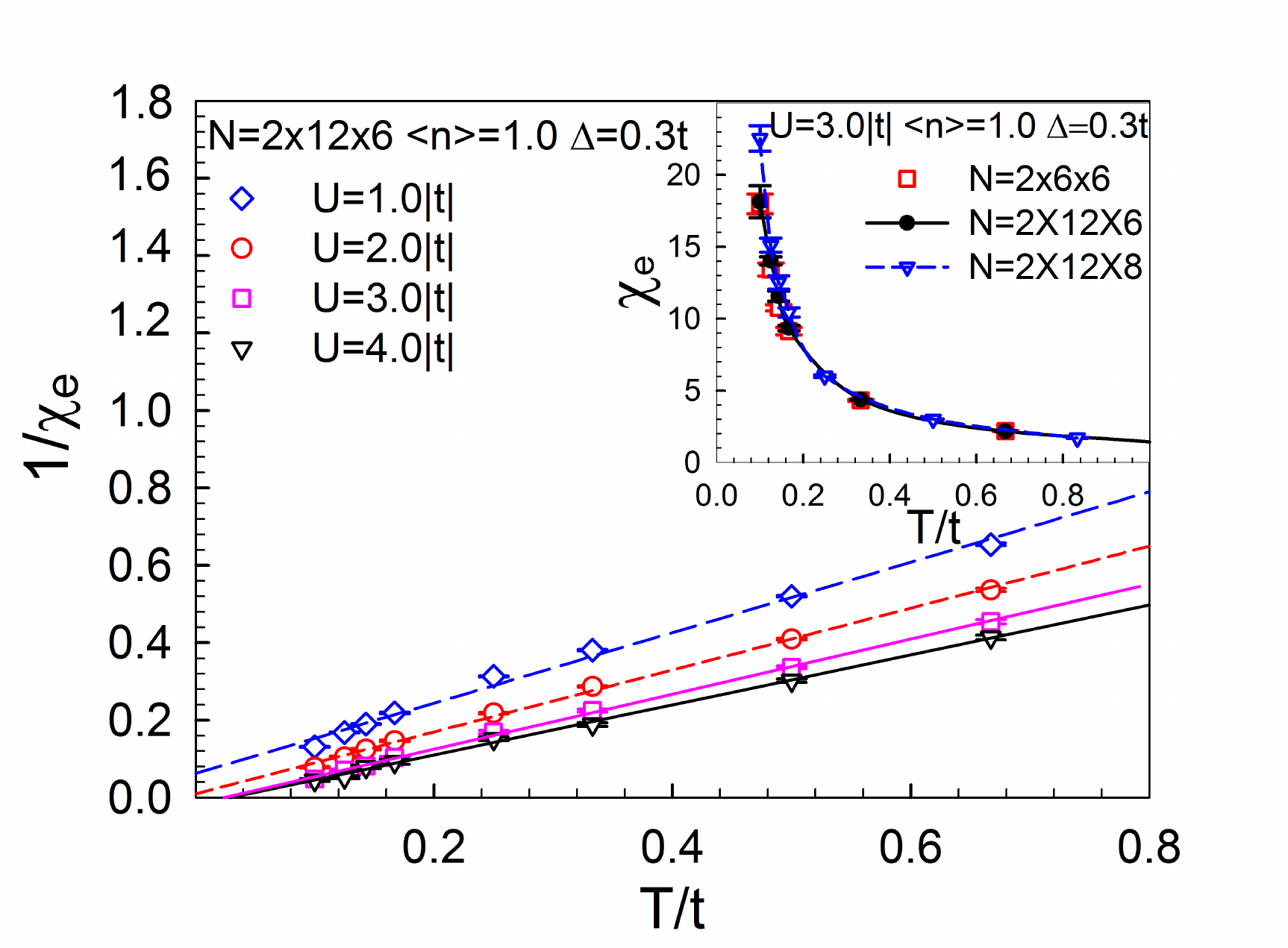}
\caption{(Colour online)The edge magnetic susceptibility $\chi_{e}$ at strain $\Delta=0.3t$ with different interactions $U$.}
\label{Fig:3}
\end{figure}

To learn more about the physics scenarios induced by the Coulomb interaction $U$, we compute the $\chi_{e}$ of graphene nanoribbons with different Coulomb interactions $U$ at the same strain in Fig.\ref{Fig:3}, and the fitting data are also shown. It is clear that for the same temperature and strain, $\chi_{e}$ is enhanced by the interaction $U$. At the same time, the ferromagnetic fluctuations dominate as $U\geqslant2.0t$ for $\Delta=0.3t$. Compared with the results shown in Fig.\ref{Fig:2}, we can find that both the Coulomb interaction and strain can boost the emergence of ferromagnetic-like behaviour in graphene nanoribbons. The results indicate that edge magnetism could be much more promisingly realized in such a strained graphene nanoribbon, as the required interaction strength is very close to the value of electronic correlation in real materials.
To detect whether our conclusions obtained above are dependent on the lattice size, we study the magnetic susceptibility for different lattice sizes in the inset of Fig. \ref{Fig:3}. The results for the same width $L_y=6$ with different length, $L_x=6$ and $L_x=12$, are almost the same within the margin of error.
For the same length $L_x=12$, results with different width, $L_y=6$ and $L_y=8$, are shown as $2\times 12 \times 6$ (dark circle) and $2\times 12 \times 8$ (blue triangle).
From the shown results, the edge magnetic susceptibility $\chi_{e}$ is slightly strengthened at low temperature as the width increases.

\begin{figure}
\centering
\includegraphics[scale=0.38]{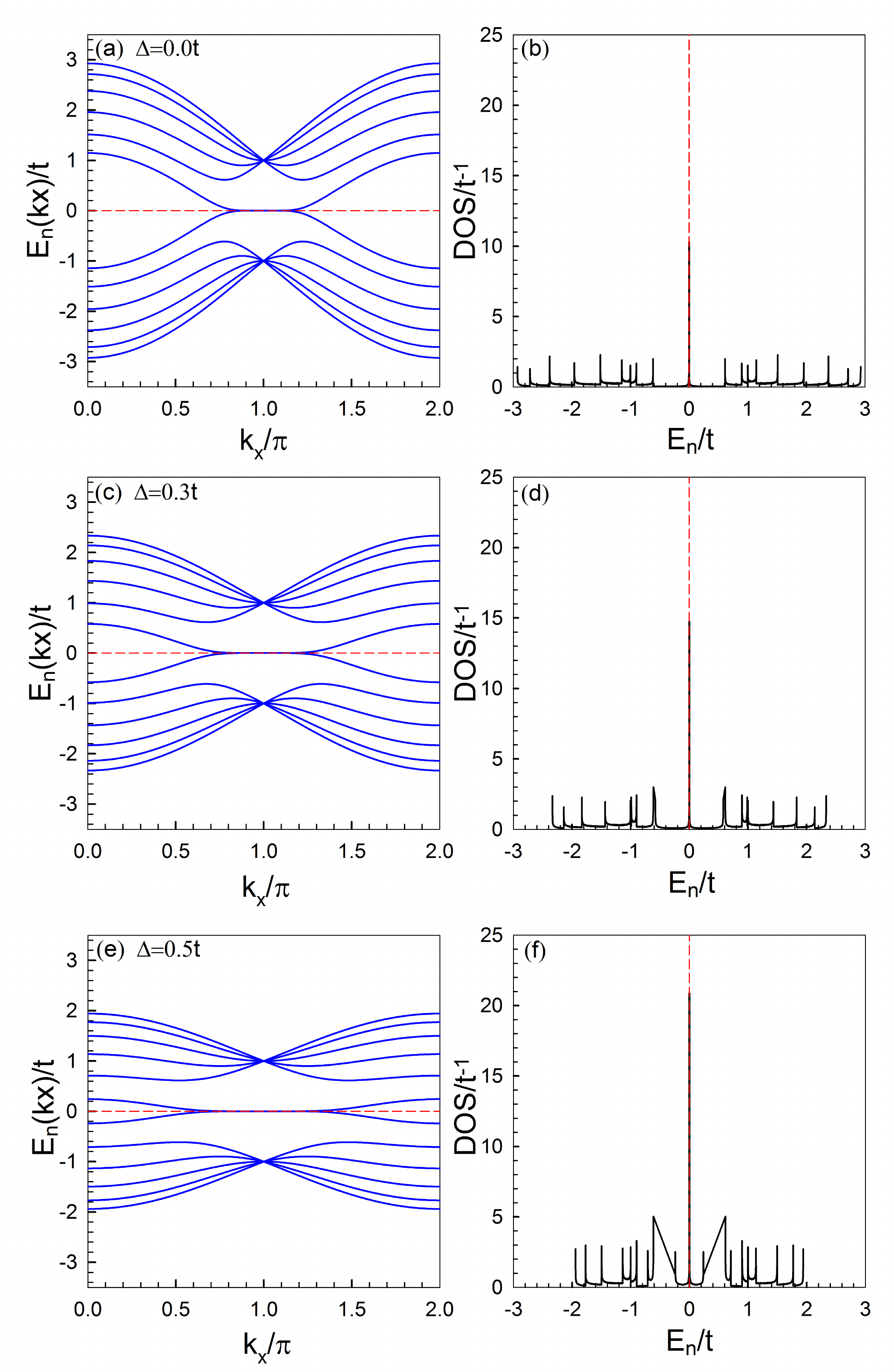}
\caption{(Colour online) Band structure((a),(c),(e)) and density of states(DOS)((b),(d),(f)) of a 6-chain graphene nanoribbon system. There is the flat band located at Fermi level in (a),(c),(e) which give rise to the DOS peak in (b),(d),(f). The red dashed lines of all figures correspond to the Fermi level of the nanoribbon system at half fillings.}
\label{Fig:Dos}
\end{figure}


The enhanced edge magnetism in such systems can be understood from the change of the topological structure induced by the strain. For the isotropic case of graphene, the length of the edge flat band in the one-dimensional Brillouin zone is $2\pi/3$, and ferromagnetism develops in these edge flat bands due to the enhanced interaction effect, indicated by the divergence in the density of states. The strain leads to a strong anisotropy in the hopping parameters, and the edge flat band may extend further over the Brillouin zone,
which leads to stronger ferromagnetism.  As is shown in Fig.\ref{Fig:Dos} (a),(c) and (e), the flat band near the Fermi level of the half filled system extend
 as the strain increases, and such an extended flat band bottom leads to a stronger peak in the density of states at half filling, as that illustrated in Fig.\ref{Fig:Dos} (b), (d) and (f). The decrease of hopping parameters caused by strain also indicates that the effective interaction is increased and thus that the required interaction strength for ferromagnetism is lowered.

\begin{figure}
\centering
\includegraphics[scale=0.5]{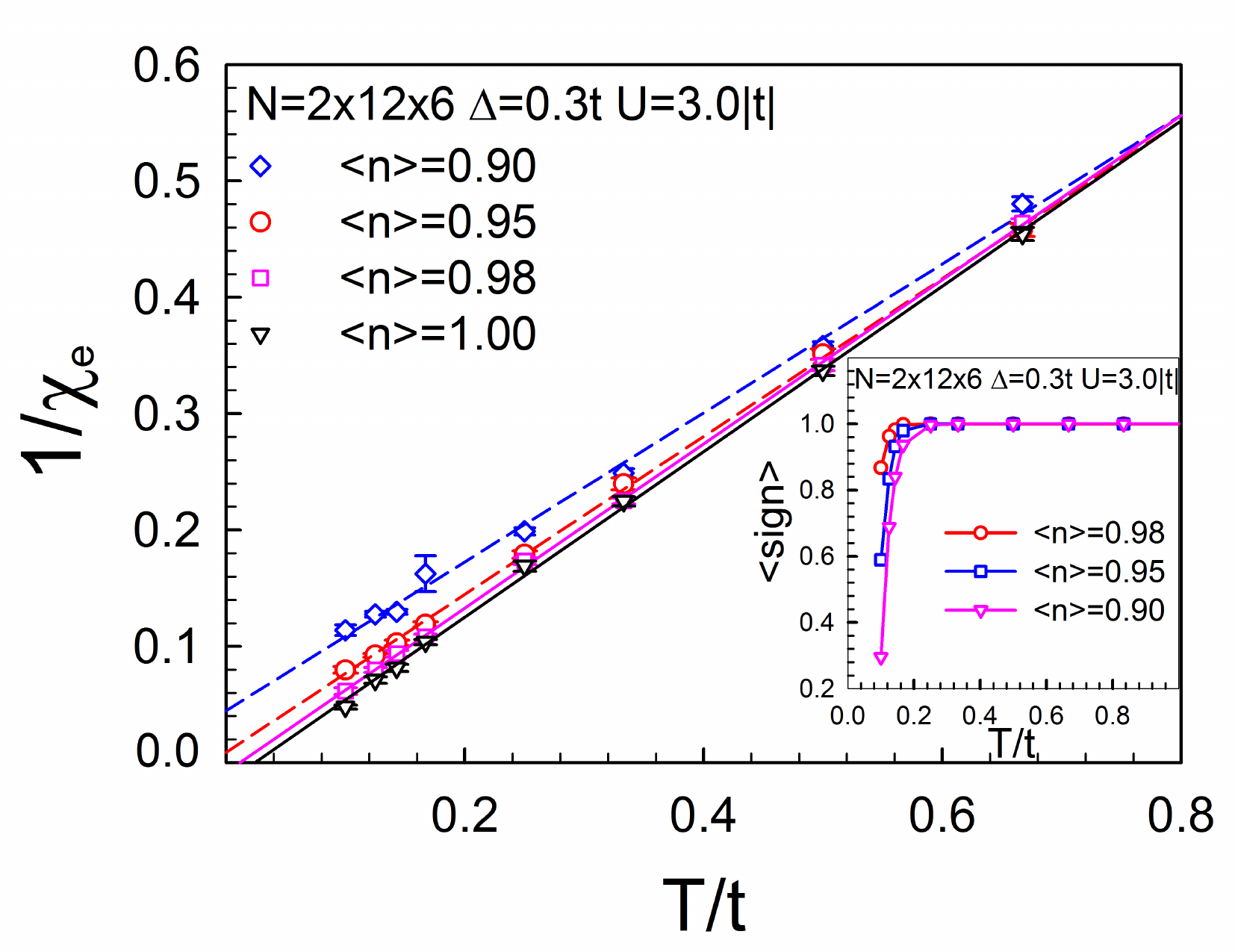}
\caption{(Colour online) The edge magnetic susceptibility at $U=3.0t$ and $\Delta=0.3t$ for different fillings.}
\label{Fig:doping}
\end{figure}

In graphene-based materials, one key issue is that their chemical potential can be tuned by an external electric field, which means that the electron filling can be changed. Thus, the filling-dependent magnetism should also be an interesting point of investigation in doped graphene nanoribbons. In Fig.\ref{Fig:doping}, the $\chi_{e}$ of a graphene nanoribbon with different electron fillings at the same strain and Coulomb interaction strength is shown. One can see that, as the electron filling moves further from half filling, the magnetism tends to be weakened, and the ferromagnetic-like behaviour is suppressed when the doping is larger than $5$ percent.

 As the system is doped away from the half filling, the particle-hole symmetry is broken and the finite
 temperature quantum Monte Carlo method serves the notorious sign problem, which prevents exact results for lower temperature, higher interaction, or larger lattice.
To make sure the data present in Fig. \ref{Fig:doping} are reliable,
the average of sign depending on temperature $T$ at different electron fillings are shown in the inset of Fig. \ref{Fig:doping},
with the Monte Carlo parameters of 30 000 times runs.
For the results near half filling, our numerical results are reliable as one can see that the average of
corresponding sign is mostly larger than 0.90 for  $U=3.0|t|$ and $\Delta t=0.3t$ with 30 000 times measurements. For electron
fillings away from the half filling, the average of sign decreases as the temperature is lowering, while
it is larger than 0.25 for the lowest temperature we reached.
In order to obtain the same quality of data as $\langle sign\rangle\simeq 1$,
the Monte Carlo runs has been stretched by a factor on the order of is $\langle sign\rangle^{-2}$\cite{PhysRevD.24.2278}.  In our simulations, some of the results are obtained with more than 500 0000 times runs, and thus the results for the current parameters are
reliable.

\begin{figure}
\centering
\includegraphics[scale=0.4]{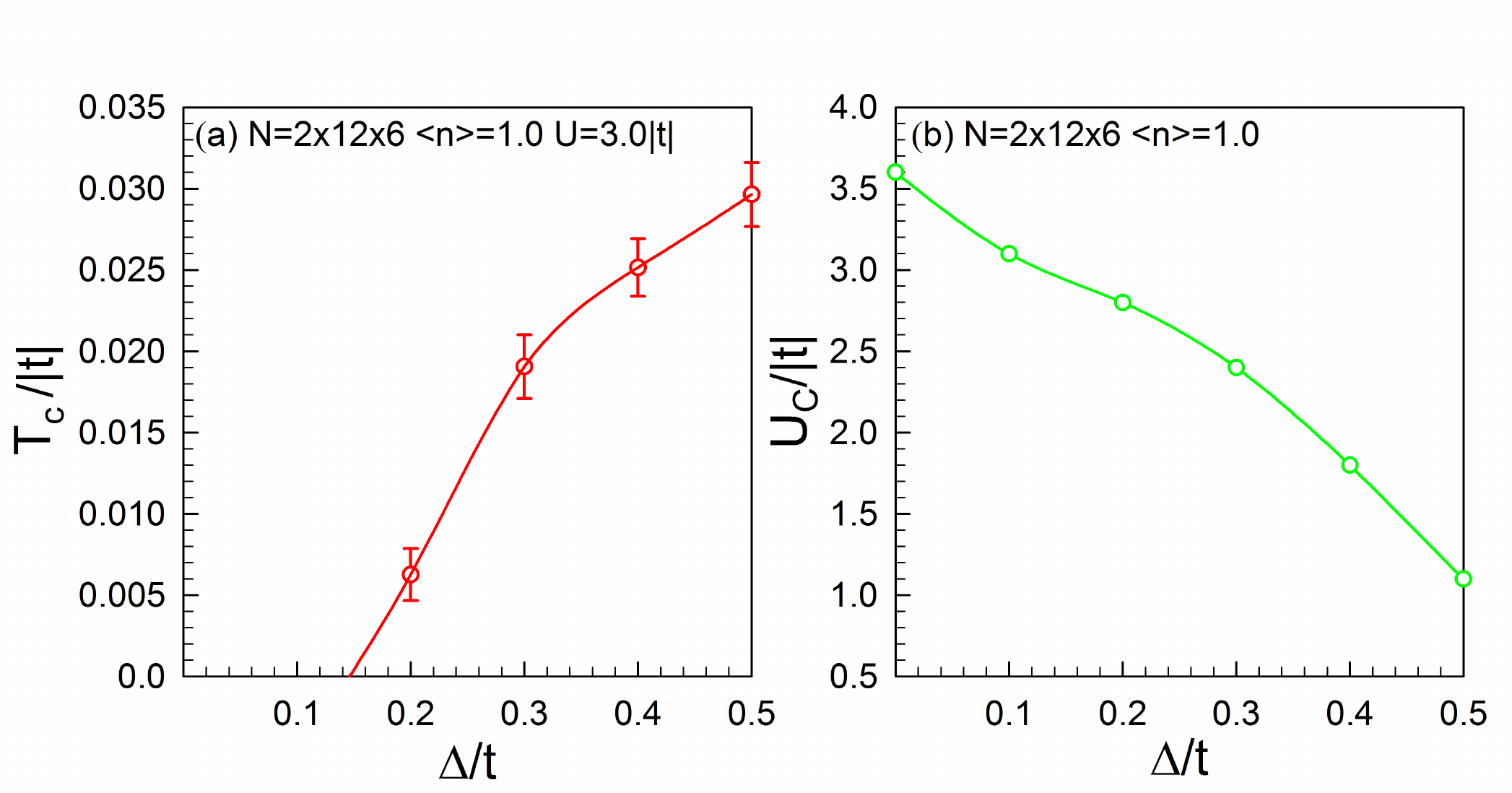}
\caption{ (Colour online)(a) Transition temperature $T_c$ for different strains and (b) the critical value of $U$ for different strains.}
\label{Fig:Uc}
\end{figure}

It has been argued that there might be room temperature magnetic order on zigzag edges of narrow graphene nanoribbons with the proper interaction strength\cite{Son2006}. In Fig. \ref{Fig:Uc}(a), we show that the temperature required for the magnetic order can be increased by increasing the strain. More importantly,  in Fig. \ref{Fig:Uc}(b), we plot the critical interaction $U_c$ as a function of strain, in which one can see that the required $U_c$ for magnetism decreases as the strain increases. At $\Delta=0.40t$, the required $U_c$ is approximately $1.6t$, and for $\Delta=0.50t$, the possible required $U_c$ is approximately $0.70t$, which means that it is very promising to realize room temperature magnetic order in strained graphene nanoribbons, as both the required interaction strength and strain are realistic in experiments.  


\section{Conclusion}

In summary, we have studied the ferromagnetic properties in strained zigzag graphene nanoribbons by using the determinant quantum Monte-Carlo method. We found that the edge magnetic susceptibility $\chi_{e}$ is enhanced greatly by the strain, and as a result, the critical interaction $U_{c}$ for magnetism at a given temperature is reduced by the strain. Our proposal, based on a nonperturbative numerical method, provides another way to tune the edge magnetism even at room temperature in zigzag graphene nanoribbons, which may be helpful to spintronics and many other applications.

{\it Acknowledgement}: We thank W. M. Sun and Y. Sun for discussions and performing simulations at the first stage. This work is supported in part by NSCFs (grant nos. 11374034 and 11334012) and the Fundamental Research Funds for the Center Universities, grant no. ~2014KJJCB26. We also acknowledge computational support from the Beijing Computational Science Research Center (CSRC), support from the HSCC of Beijing Normal University, and the Special Program for Applied Research on Super Computation of the NSFC-Guangdong Joint Fund (the second phase). M. Ye acknowledges support from NSCF (no. 61662018) and the Guangxi Natural Science Foundation of China (no. 2016GXNSFAA380153).

\bibliography{reference}
\end{document}